\documentclass[twocolumn,showpacs,prl]{revtex4}

\usepackage{amsmath,amssymb}
\usepackage{graphicx}

\draft

\begin{document}

\title{\large \bf Quantum Memory Process with a Four-Level Atomic Ensemble}

\author{Xiong-Jun Liu$^{a,b,c}$\footnote{Electronic address: phylx@nus.edu.sg}, Hui Jing$^d$ and Mo-Lin Ge$^{b,c}$}

\affiliation{a. Department of Physics, National University of
Singapore,
2 Science Drive 3, Singapore 117542 \\
b. Theoretical Physics Division, Nankai Institute of
Mathematics,Nankai University, Tianjin 300071, P.R.China\\
c. Liuhui Center for Applied Mathematics, Nankai
University and Tianjin University, Tianjin 300071, P.R.China\\
 d. State Key Laboratory of Magnetic Resonance and Atomic and Molecular
 Physics,\\
 Wuhan Institute of Physics and Mathematics, CAS, Wuhan 430071, P. R. China}

\begin{abstract}
We examine in detail the quantum memory technique for photons in a
double $\Lambda$ atomic ensemble in this work. The novel
application of the present technique to create two different
quantum probe fields as well as entangled states of them is
proposed. A larger zero-degeneracy class besides dark-state
subspace is investigated and the adiabatic condition is confirmed
in the present model. We extend the single-mode quantum memory
technique to the case with multi-mode probe fields, and reveal the
exact pulse matching phenomenon between two quantized pulses in
the present system.\\

\end{abstract}

\maketitle

\section{introduction}

\indent Since the remarkable demonstration of ultraslow light
speed in a Bose-Einstein condensate in 1999 \cite{1}, rapid
advances have been witnessed in both experimental and theoretical
aspects towards probing the novel mechanism of Electromagnetically
Induced Transparency (EIT) \cite{2} and its many potential
applications \cite{3,4,5,entangled,wu}. Particularly, based on
``dark-state polaritons" (DSPs) theory \cite{6}, the quantum
memory via EIT technique is actively being explored by
transferring the quantum states of photon wave-packets to
metastable collective atomic-coherence (collective quasi spin
states) in a loss-free and reversible manner \cite{7}. For the
three-level EIT quantum memory technique, a semidirect product
group under the condition of large atom number and low collective
excitation limit \cite{6} was discovered by Sun $et al.$ \cite{8},
and the validity of the adiabatic condition for the evolution of
DSPs has also been confirmed.

As a natural extension, controlled light storing in a medium
composed of double $\Lambda$ type four-level atoms was mentioned
\cite{9} and briefly studied recently \cite{10}. However, in these
previous theoretical works, the probe light is treated as
classical\cite{10} and the evolution of the total wave function of
the probe pulses and atoms is not clear. Thus many properties of
quantum memory with four-level atomic system have not been
discovered. In this paper, we present a quantum description of DSP
theory in such a double $\Lambda$ type atomic ensemble interacting
with two quantized fields and two classical control fields. The
novel application of our model to create two different quantum
probe fields as well as their entangled states is proposed.
Furthermore, we extend the single-mode quantum memory technique to
the case with multi-mode probe fields, and reveal the exact pulse
matching phenomenon between two quantized probe pulses.

\section{model}
\begin{figure}[ht]
\includegraphics[width=0.9\columnwidth]{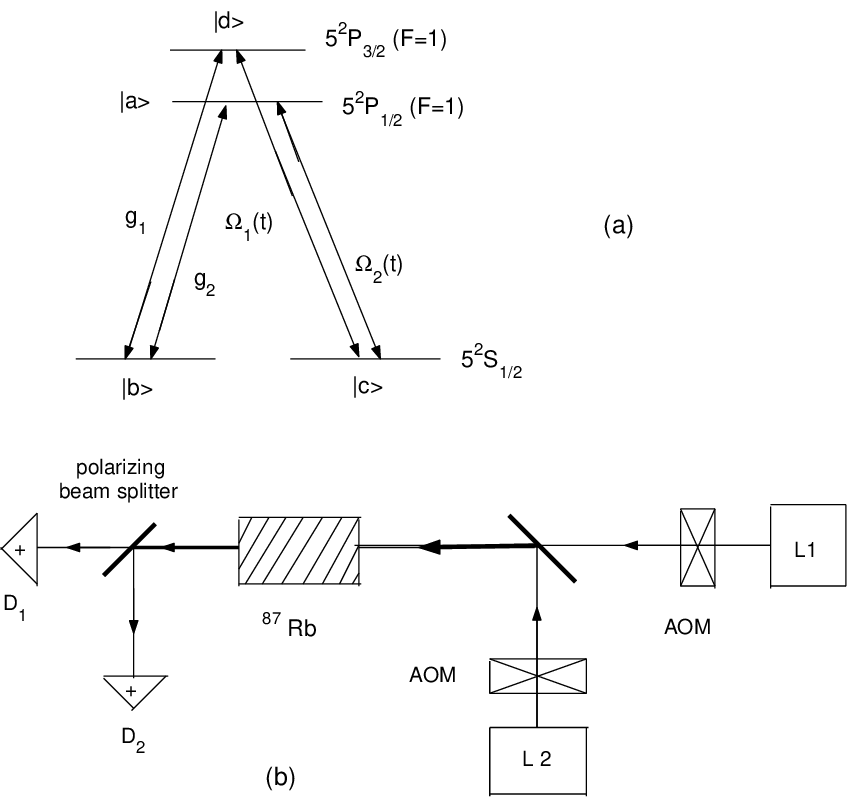}
\caption{Double $\Lambda$ type four-level $^{87}$Rb atoms coupled
to two single-mode quantized and two classical control fields (a).
The schematic setup for experimental realization is shown in part
(b). Co-propagating input probe and control fields are used to
avoid Doppler-broadening. The polarizing beam splitter is used to
separate probe photons from control ones.}\label{1}
\end{figure}
Turning to the situation of Fig. 1(a), we assume that a collection
of $N$ double $\Lambda$ type four-level atoms ($^{87}$Rb) interact
with two single-mode quantized fields with coupling constants
$g_1$ and $g_2$, and two classical control ones with
time-dependent real Rabi-frequencies $\Omega_1(t)$ and
$\Omega_2(t)$. Generalization to the multi-mode probe pulse case
will be studied later. All probe and control fields are
co-propagating in the $z$ direction (Fig. 1(b)). Considering all
transitions at resonance, the interaction Hamiltonian of the total
system can be written as:
\begin{eqnarray}\label{eqn:1}
\hat V=g_1\sqrt{N}\hat a_1\hat A^{\dag}+\Omega_1\hat
T_{ac}+g_2\sqrt{N}\hat a_2\hat D^{\dag}+\Omega_2\hat T_{dc}+h.c.,
\end{eqnarray}
where the collective atomic excitation operators: $\hat
A=\frac{1}{\sqrt{N}}\sum_{j=1}^{N}\hat\sigma_{ba}^{j}, \ \hat
C=\frac{1}{\sqrt{N}}\sum_{j=1}^N\hat\sigma_{bc}^{j}, \ \hat
D=\frac {1}{\sqrt{N}}\sum_{j=1}^N\hat\sigma _{bd}^{j}$ with
$\hat\sigma^i_{\mu\nu}=|\mu\rangle_{ii}\langle\nu|
(\mu,\nu=a,b,c,d)$ being the flip operators of the $i$-th atom
between states $|\mu\rangle$ and $|\nu\rangle$, and $\hat
T^{-}_{\mu\nu}=\hat T_{\mu\nu}=\sum_{j=1}^{N}\hat\sigma
_{\mu\nu}^{j}, \ \ \hat T_{\mu\nu}^{+}=(\hat
T^{-}_{\mu\nu})^{\dagger }$ with $\mu\neq\nu=a,c,d$. Denoting by
$|b\rangle=|b_1,b_2,...,b_N\rangle$ the collective ground state
with all $N$ atoms staying in the same single particle ground
state $|b\rangle$, we can easily give other quasi-spin wave states
by the collective atomic excitation operators:
$|a^n\rangle=[n!]^{-1/2}(\hat A^{\dag})^n|b\rangle$,
$|c^n\rangle=[n!]^{-1/2}(\hat C^{\dag})^n|b\rangle$, and
$|d^n\rangle=[n!]^{-1/2}(\hat D^{\dag})^n|b\rangle$. Following the
analysis in ref. \cite{8}, one can verify that the dynamical
symmetry of our double $\Lambda$ system is governed by a
semidirect sum Lie algebra $su(3)\overline{\otimes}h_3$ in large
$N$ limit and low excitation condition.

To give a clear description of the interesting quantum memory
process in this double $\Lambda$ type four-level-atoms ensemble,
we define the new type of dark-state-polaritons operator as
\begin{equation}\label{eqn:6}
\hat d=\cos\theta\cos\phi\hat a_1-\sin\theta \hat
C+\cos\theta\sin\phi\hat a_2,
\end{equation}
where the mixing angles $\theta$ and $\phi$ are defined through
$\tan\theta=g_1\sqrt{N}/\sqrt{\Omega_1^2+\Omega_2^2g_1^2/g_2^2}$
and $\tan\phi=g_1\Omega_2/g_2\Omega_1$. By a straightforward
calculation one can verify that $[\hat d,\hat d^{\dag}]=1$ and $
[\hat V,\hat d \ ]=0 $, hence the general atomic dark states can
be obtained through $|D_n\rangle=[n!]^{-1/2}(\hat
d^{\dag})^n|0\rangle$, where
$|0\rangle=|b\rangle\otimes|0,0\rangle_e$ and $|0,0\rangle_e$
denotes the electromagnetic vacuum of two quantized probe fields.
So we reach
\begin{eqnarray}\label{eqn:8}
|D_n\rangle&=&\sum^n_{k=0}\sum^{n-k}_{j=0}\sqrt{\frac{n!}{k!(n-k-j)!j!}}(-\sin\theta)^k(\cos\theta)^{n-k}\nonumber\\
&&(\sin\phi)^j(\cos\phi)^{n-k-j}|c^k,n-k-j,j\rangle .
\end{eqnarray}
From this formula it is clear that when the mixing angle $\theta$
is adiabatically rotated from $0$ to $\pi/2$, the quantum state of
the DSPs is transferred from pure photonic character to collective
excitations, i.e. $|D_n\rangle: \
\sum^{n}_{j=0}\sqrt{\frac{n!}{(n-j)!j!}}
(\sin\phi)^j(\cos\phi)^{n-j}|b\rangle|n-j,j\rangle\rightarrow
|c^n\rangle|0,0\rangle$.

Similarly, another important physical phenomenon can also be
predicted through our quantized description of this system. If
initially only one quantized field (described by the coherent
state $|\alpha_2\rangle$ with $\alpha_2=\alpha_0$) is injected
into the atomic ensemble to couple the transition from $|b\rangle$
to $|d\rangle$, and the second control field is chosen to be much
stronger than the first one $(g_1\Omega_2(0)\gg g_2\Omega_1(0))$
along with $\sqrt{g_2^2\Omega_1^2(0)+g_1^2\Omega_2(0)^2}\gg
g_1g_2\sqrt{N}$ (or $\sin\phi_0=1$ and $\cos\theta_0=1$), the
initial total state of the quantized field and atomic ensemble
reads
\begin{eqnarray}\label{eqn:9}
|\Psi_0\rangle=\sum_{n}P_n(\alpha_0)|0,n\rangle\otimes|b\rangle ,
\end{eqnarray}
where
$P_n(\alpha_0)=\frac{\alpha_0^n}{\sqrt{n!}}e^{-|\alpha_0|^2/2}$ is
the probability distribution function. Subsequently, the mixing
angle $\theta$ is adiabatically rotated to $\pi/2$ by turning off
the two control fields, hence the quantum state of the probe light
$|\alpha_2\rangle$ is fully mapped into the collective atomic
excitations. When both of the two control fields are turned back
on and the mixing angle $\theta$ is rotated to $\theta=0$ again
with $\phi$ to some value $\phi_e$, which is only determined by
the Rabi-frequencies of the two re-applied control fields, we
finally obtain from eq.(\ref{eqn:8})
\begin{eqnarray}\label{eqn:10}
|\Psi_e\rangle&=&\sum_{j}\sum_{k}P_j(\alpha_{e1})P_k(\alpha_{e2})
|b,j,k\rangle\nonumber\\
&&=|b\rangle\otimes|\alpha_{e1}\rangle\otimes|\alpha_{e2}\rangle,
\end{eqnarray}
where $\alpha_{e1}=\alpha_0\cos\phi_e$ and
$\alpha_{e2}=\alpha_0\sin\phi_e$ are the parameters of the two
released coherent lights. The above expression shows that the
injected quantized field can convert into two different coherent
pulses $|\alpha_{ei}\rangle (i=1,2)$ after a proper evolution
manipulated by two control fields. For example, i) if
$\phi_e=\pi/2$, we have $\alpha_{e1}=0$ and
$\alpha_{e2}=\alpha_0$, which means the released pulse is the same
as the initial injected one; ii) if $\phi_e=0$, we have
$\alpha_{e1}=\alpha_0$ and $\alpha_{e2}=0$, which means that the
injected quantized field state is now fully converted into a
different light beam $|\alpha_{e1}\rangle$. Obviously, this novel
mechanism can be extended to other cases of the injected field,
for example, in the presence of a non-classical or squeezed light
beam (see the following discussion). In experiments, there also
holds the promise for actual observation through, e.g., combining
a beam splitter and an electro-optic modulator to generate the
requisite sidebands \cite{1}.

\section{generation of entangled coherent states}
It is interesting to note that when we input a non-classical or
squeezed probe light, by proper steering the control fields, we
can generate two output entangled light beams. Firstly we consider
that the injected quantized field is in a macroscopic quantum
superposition of coherent states, e.g. for the initial total state
\begin{eqnarray}\label{eqn:initial1}
|\Psi_0\rangle^{\pm}=\frac{1}{\sqrt{{\cal
N}(\alpha_0)}}|0\rangle\otimes(|\alpha_0\rangle\pm|-\alpha_0\rangle)
\otimes|b\rangle,
\end{eqnarray}
where the normalized factor ${\cal
N}_{\pm}(\alpha_0)=2\pm2e^{-2|\alpha_0|^2}$, with the scheme
discussed above and from eq. (\ref{eqn:8}) we find the injected
quantized pulse can evolve into a very interesting entangled
coherent state (ECS) of two output fields
($|\Psi_0\rangle^{\pm}\rightarrow|\Psi_e\rangle^{\pm}$)
\begin{widetext}
\begin{eqnarray}\label{eqn:entangled}
&\frac{1}{\sqrt{{\cal
N}_{\pm}(\alpha_0)}}|0\rangle\otimes\bigr(|\alpha_0\rangle\pm|-\alpha_0\rangle\bigr)
\otimes|b\rangle=\frac{1}{\sqrt{{\cal
N}_{\pm}(\alpha_0)}}\bigr(|0\rangle\otimes|\alpha_0\rangle\pm|0\rangle\otimes|-\alpha_0\rangle\bigr)
\otimes|b\rangle\longrightarrow\nonumber\\
&\longrightarrow\frac{1}{\sqrt{{\cal
N}_{\pm}(\alpha_0)}}\bigr(\sum_{j}\sum_{k}P_j(\alpha_{e1})P_k(\alpha_{e2})
|b,j,k\rangle\pm\sum_{j}\sum_{k}P_j(-\alpha_{e1})P_k(-\alpha_{e2})
|b,j,k\rangle\bigr).
\end{eqnarray}
The final state in the above formula can be rewritten as:
\begin{eqnarray}\label{eqn:entangled1}
|\Psi_e\rangle^{\pm}=\frac{1}{\sqrt{{\cal
N}_{\pm}(\alpha_0)}}\bigr(|\alpha_{e1},\alpha_{e2}
\rangle\pm|-\alpha_{e1},-\alpha_{e2}\rangle\bigr)_{light}
\otimes|b\rangle.
\end{eqnarray}
\end{widetext}
where the subscript $light$ indicates the state of the output two
probe pulses.

If $\phi_e=0$, hence $\alpha_{e1}=\alpha_0$ and $\alpha_{e2}=0$,
and then the evolution of the quantized fields proceeds as
$|0\rangle\otimes(|\alpha_0\rangle\pm|-\alpha_0\rangle)/\sqrt{{\cal
N}_{\pm}(\alpha_0)}\rightarrow(|\alpha_0\rangle\pm|-\alpha_0\rangle)\otimes|0\rangle/\sqrt{{\cal
N}_{\pm}(\alpha_0)}$, which means the input Schr\"{o}dinger state
is now fully converted into another one with different vibrational
mode. On the other hand, for the general case of non-zero value of
the coherent parameters $\alpha_{e1}$ and $\alpha_{e2}$, the
states of output quantized fields are entangled coherent states.
Since the parameters $\alpha_{ei} (i=1,2)$ are controllable, the
entanglement of the output states \cite{entanglement}
$E^{\pm}(\alpha_{e1}, \alpha_{e2})=-$
tr$(\rho^{\pm}_{\alpha_{e1}}\ln\rho^{\pm}_{\alpha_{e1}})$ with the
reduced density matrix $\rho^{\pm}_{\alpha_{e1}}=$
tr$^{(\alpha_{e2}, atom)}(|\Psi_e\rangle\langle\Psi_e|)^{\pm}$ can
also easily be controlled by the re-applied control fields. In
particular, for the initial state $|\Psi_0\rangle^{-}$, if
$\phi_e=\pi/4$, we have
$\alpha_{e1}=\alpha_{e2}=\alpha_0/\sqrt{2}$ and then we obtain the
maximally entangled state(MES):
\begin{eqnarray}\label{eqn:entangled2}
&&|0\rangle\otimes\bigr(|\alpha_0\rangle-|-\alpha_0\rangle\bigr)/\sqrt{{\cal
N}_-(\alpha_0)}\longrightarrow\nonumber\\
&&\longrightarrow\bigr(|\frac{\alpha_0}{\sqrt{2}},\frac{\alpha_0}
{\sqrt{2}}\rangle-|-\frac{\alpha_0}{\sqrt{2}},-\frac{\alpha_0}{\sqrt{2}}\rangle\bigr)/\sqrt{{\cal
N}_-(\alpha_0)},
\end{eqnarray}
which is most useful for quantum information processes. With the
definitions of the orthogonal basis
$|+\rangle=\bigr(|\frac{\alpha_0}{\sqrt{2}}\rangle+|-\frac{\alpha_0}{\sqrt{2}}\rangle\bigr)/\sqrt{{\cal
N}_+(\alpha_0/2)}$ and
$|-\rangle=\bigr(|\frac{\alpha_0}{\sqrt{2}}\rangle-|-\frac{\alpha_0}{\sqrt{2}}\rangle\bigr)/\sqrt{{\cal
N}_-(\alpha_0/2)}$, the output coherent states can be rewritten as
\begin{eqnarray}\label{eqn:entangled3}
\Phi_{light}(-)=\frac{1}{\sqrt{2}}\bigr(|+\rangle|-\rangle+|-\rangle|+\rangle\bigr)_{light},
\end{eqnarray}
which manifestly has one ebit of entanglement (since
$\langle+|-\rangle=0$). We should emphasize that all the above
results can not be obtained with classical DSP theory of a
four-level system. Since our scheme of generating the entangled
coherent states via quantized DSP theory is linear and
controllable and it only requires a macroscopic quantum
superposition for the initial state, this scheme may be feasible
in experiment which has made much progress in recent years
\cite{ent}. Besides our technique, the generation of entangled
coherent states via Kerr effect \cite{entangled} and entanglement
swapping using Bell-state measurements \cite{swap} is also studied
widely.

If the two output entangled coherent lights are respectively
injected into two other atomic ensembles composed of many
three-level atoms, and the quantum states of the lights are mapped
into quasi spin-waves via sperate Raman transitions, it is
possible to generate controllable entangled coherence of two
atomic ensembles.

Consider now a different type of input quantum state corresponding
to a single-photon state, i.e. the initial total state
$|\Psi_0\rangle=(|0\rangle\otimes|1\rangle)_{light}\otimes|b\rangle$.
According to Eq. (\ref{eqn:8}) and after the light state storage
process discussed in section II, the final state yields:
\begin{eqnarray}\label{eqn:entangled3}
\Phi_{light}=\frac{1}{\sqrt{2}}\bigr(|1\rangle|0\rangle+|0\rangle|1\rangle\bigr)_{light}.
\end{eqnarray}

The entangled states generated with the present scheme have other
interesting aspects. Firstly, since the two output probe fields
are different in frequency, the generated entangled states is
between two quantized fields with different frequencies. Secondly,
since the direction of the output probe field can be fully
controlled by the corresponding control field \cite{3}, based on
our scheme, the output directions of the two entangled probe
fields can be controlled by the two reapplied control fields.
These interesting factors are advantages of our scheme for
generating entangled light fields, which is different from that
using a standard beam splitter.

\section{validity of adiabatic condition}
As we have known, the condition of adiabatic evolution is most
important for the quantum memory technique based on the quantized
DSPs theory, because the total system should be confined to the
dark states during the process of quantum memory. One can verify
that when $g_1\neq g_2$, no larger zero subspace is obtained
except for dark states and the adiabatic condition can be
guaranteed by the adiabatic theorem. However, the dynamical
symmetry in the present system depicted by the semi-direct sum
algebra $h_3\overline{\otimes}su(3)$ indicates that, for the
special case $g_1=g_2=g$, we may find a larger degeneracy class of
states with zero-eigenvalue in this system. We define
\begin{eqnarray}\label{eqn:operator1}
\hat Q_{\pm}^{\dag}=\hat u^{\dag}\pm\hat b^{\dag}, \ \ \hat
P_{\pm}^{\dag}=-\sin\phi\hat a_1+\cos\phi\hat a_2\pm\hat v^{\dag},
\end{eqnarray}
where the operators $\hat u$, $\hat v$ and the
bright-state-polaritons (BSPs) operator $\hat b$ are defined as:
$\hat u=\cos\phi\hat A+\sin\phi\hat D , \
 \hat v=-\sin\phi\hat A+\cos\phi\hat D$ and $\hat b
=\sin\theta\cos\phi\hat a_1+\cos\phi\hat C +\sin\theta\sin\phi\hat
a_2$. By a straightforward calculation one obtains the
communication relations $[\hat V,\hat
Q_{\pm}^{\dag}]=\pm\epsilon_1\hat Q_{\pm}^{\dag}$, $[\hat V,\hat
P^{\dag}]=\pm\epsilon_2\hat P_{\pm}^{\dag}$ and $[\hat
P_{\pm}^{\dag},\hat Q_{\pm}^{\dag}]=0$ with
$\epsilon_1=\sqrt{g^2N+\Omega_1^2+\Omega_2^2}$ and
$\epsilon_2=g\sqrt{N}$. Thus we further obtain
\begin{eqnarray}\label{eqn:com}
[\hat V,\hat P_{\pm}^{\dag}\hat Q_{\pm}^{\dag}]=\pm\epsilon_1\hat
P_{\pm}^{\dag}\hat Q_{\pm}^{\dag} \pm \epsilon_2\hat
P_{\pm}^{\dag}\hat Q_{\pm}^{\dag}
\end{eqnarray}
To this end we have obtained all communication relations between
the above operators. Thanks to these results we finally obtain a
much larger degeneracy class:
\begin{eqnarray}\label{eqn:degen}
|r(i,j;k,l;n)\rangle=\frac{1}{\sqrt{i!j!k!l!}}(\hat
Q_+^{\dag})^i(\hat Q_-^{\dag})^j(\hat P_+^{\dag})^k(\hat
P_-^{\dag})^l|D_n\rangle\nonumber,
\end{eqnarray}
with eigenvalue $E(i,j;k,l)=(i-j)\epsilon_1+(k-l)\epsilon_2$.
Obviously, when $i=j$ and $k=l$, one finds the zero-eigenvalue
degeneracy class is
\begin{eqnarray}\label{eqn:11}
|d(i,k;n)\rangle&=&\frac{1}{i!k!}(\hat Q_+^{\dag}\hat
Q_-^{\dag})^i(\hat P_+^{\dag}\hat P_-^{\dag})^k|D_n\rangle,\\
&&(i,k,n=0,1,2,\cdots)\nonumber.
\end{eqnarray}

The larger class $\{|d(i,k;n)\rangle \ |$ $\ n=0,1,2,\cdots \}$ of
states of zero eigenvalue are constructed by acting ($\hat
Q_+^{\dag}\hat Q_-^{\dag}$) $i$ times and ($\hat P_+^{\dag}\hat
P_-^{\dag}$) $k$ times on the dark state $|D_n\rangle$. Only when
$i=0$ and $k=0$, the larger degeneracy
class reduces to the special subset $%
\{|D_n\rangle \ |$$ n=0,1,2,\cdots \}$ of the interaction
Hamiltonian. As usual, the quantum adiabatic theorem does not
forbid the transition between those states of same eigenvalue,
hence it is important also in the present four-level-atoms system
to confirm the forbiddance of any transitions from dark states
$|D_n\rangle$ to $\{ |d(i,k;n)\rangle \ |ik\neq0,n=0,1,2,\cdots
\}$. Generally this problem can be studied by defining the zero-eigenvalue subspaces $%
{\bf S}^{[i,k]}:\{|d(i,k;n)\rangle \ |i,k,n=0,1,2,\cdots \}$, in which ${\bf S}^{[0,0]}={\bf S}%
$ is the dark-state subspace. The complementary part of the direct sum
${\bf DS}={\bf S}^{[0,0]}\oplus $ $%
{\bf S}^{[0,1]}\oplus$ $%
{\bf S}^{[1,0]}\oplus\cdots$ of all zero-eigenvalue subspaces is
noted by ${\bf ES}={\bf S}^{[\bf ES]}$ in which each state turns
out to have some nonzero eigenvalue after some calculations. Any
state $|\phi
^{\lbrack i,k]}(t)\rangle =\sum_{i,k;n}c_{n}^{[i,k]}(t)|d(i,k;n)\rangle $ in ${\bf S}%
^{[i,k]}$ evolves according to \cite{8}
\begin{equation}\label{eqn:14}
i\frac{d}{dt}c_{n}^{[i,k]}(t)=\sum_{i^{\prime },k^{\prime
};n^{\prime }}D_{i,k;n}^{i^{\prime },k^{\prime };n^{\prime
}}c_{n^{\prime }}^{[i^{\prime },k^{\prime }]}(t)+F[{\bf ES}],
\end{equation}
where $F[{\bf ES}]$, which can be ignored under adiabatic
conditions \cite{8,11}, represents a certain functional of the
complementary states and $D_{i,k;n}^{i^{\prime },k^{\prime
};n^{\prime }}=-i\langle d(i^{\prime },k^{\prime };n^{\prime
})|\partial _{t}|d(i,k;n)\rangle=-i\dot{\theta}\langle d(i^{\prime
},k^{\prime };n^{\prime })|\partial
_{\theta}|d(i,k;n)\rangle-i\dot{\phi}\langle d(i^{\prime
},k^{\prime };n^{\prime })|\partial _{\phi}|d(i,k;n)\rangle $ with
$\dot{\theta}=d\theta/dt$ and $\dot{\phi}=d\phi/dt$. With the
definitions of these operators, we can easily calculate:
\begin{eqnarray}\label{eqn:15}
\partial_{\theta}\hat b=\hat d, \ \partial_{\theta}\hat d=-\hat
b;\nonumber\\
\partial_{\phi}\hat b=\sin\theta\hat s, \ \partial_{\phi}\hat
u=\hat v, \\
\partial_{\phi}\hat v=-\hat u, \
\partial_{\phi}\hat a=\hat s, \ \partial_{\phi}\hat s=-\hat a \nonumber,
\end{eqnarray}
where $\hat a= \cos\phi\hat a_1+\sin\phi\hat a_2$ and $\hat s=
-\sin\phi\hat a_1+\cos\phi\hat a_2$. From these results one can
finally determine that the equations about $\partial _{\theta
}|d(i,k;n)\rangle $ and $\partial _{\phi }|d(i,k;n)\rangle $ do
not contain the term $|d(i^{\prime },k^{\prime };n^{\prime
})\rangle$, hence $\langle d(i^{\prime },k^{\prime };n^{\prime
})|\partial _{t}|d(i,k;n)\rangle=0$ and the evolution equation
yields $\frac{d}{dt}c_{n}^{[i,k]}(t)=0$, i.e., there is no mixing
of different zero-eigenvalue subspaces during the adiabatic
process and therefore, even for the special case of $g_1=g_2$,
quantum memory may till be robust in the present double $\Lambda$
type atomic ensemble.

\section{quantum memory for multi-mode quantized fields}
In this section we shall extend the technique of quantum memory
for a single-mode field to the multi-mode case in the double
$\Lambda$ atomic-ensemble system. The two quantized fields
described by the slowly-varying dimensionless operator are given
by
\begin{equation}\label{eqn:47}
\hat{\cal E}_j(z,t) =\sum_k \hat a_{k_j}(t){\rm
e}^{-i\frac{\nu_j}{c}(z-ct)}, \ (j=1,2),
\end{equation}
where $\nu_1=\omega_{ab}, \nu_2=\omega_{db}$ are the carrier
frequencies of the two quantized optical fields. If the
(slowly-varying) quantum amplitude does not change much in a small
length interval $\Delta z$ which contains $N_z\gg 1$ atoms, we can
introduce continuous atomic variables \cite{6}
\begin{eqnarray}\label{eqn:48}
\widetilde\sigma_{\mu\nu}(z,t) =\frac{1}{N_z}\sum_{z_j\in
N_z}{\hat\sigma}_{\mu\nu}^j(t),
\end{eqnarray}
where $\hat\sigma_{\mu\nu}^j=|\mu_j\rangle\langle\nu_j| \, {\rm
e}^{-i\frac{\omega_{\mu\nu}}{c}(z-ct)}$ is the slowly-varying part
of the atomic flip operators. Making the replacement $\sum_{j=1}^N
\longrightarrow \frac{N}{L}\int {\rm d} z$ with $L$ the length of
the interaction in the propagation direction of the quantized
field, the interaction Hamiltonian then yields
\begin{eqnarray}\label{eqn:49}
{\hat V}&=&-\int\frac{dz}{L}\bigr(\hbar
g_1N\widetilde\sigma_{ab}(z,t)\hat {\cal
E}_1(z,t)+\hbar\Omega_1(t)
N\widetilde\sigma_{ac}(z,t)\nonumber\\
&&+\hbar g_2N\widetilde\sigma_{db}(z,t)\hat {\cal
E}_2(z,t)+\hbar\Omega_2(t)N\widetilde\sigma_{dc}(z,t)\\
&&+h.a\bigr).\nonumber
\end{eqnarray}

The evolution of the Heisenberg operators $\hat{\cal E}_i(z,t)$
corresponding to the two quantum fields can be described by the
propagation equations
\begin{eqnarray}\label{50}
\left(\frac{\partial}{\partial t}+c\frac{\partial}{\partial
z}\right) \hat {\cal E}_1(z,t)= ig_1N\widetilde\sigma_{ba}(z,t)
\end{eqnarray}
and
\begin{eqnarray}\label{51}
\left(\frac{\partial}{\partial t}+c\frac{\partial}{\partial
z}\right) \hat {\cal E}_2(z,t)= ig_2N\widetilde\sigma_{bd}(z,t).
\end{eqnarray}
In the condition of low excitation, i.e.
$\widetilde\sigma_{bb}\approx1$, the atomic evolution governed by
the Heisenberg-Langevin equations can be obtained by
\begin{eqnarray}\label{eqn:52}
\dot{\widetilde\sigma}_{ba}=-\gamma_{ba} {\widetilde\sigma}_{ba}
+ig_1\hat{\cal E}_1+i\Omega_1{\widetilde\sigma}_{bc} +F_{ba},
\end{eqnarray}
\begin{eqnarray}\label{eqn:53}
\dot{\widetilde\sigma}_{bc}= i\Omega_1{\widetilde\sigma}_{ba}
-ig_1\hat {\cal
E}_1{\widetilde\sigma}_{ac}+i\Omega_2{\widetilde\sigma}_{bd}
-ig_2\hat {\cal E}_2{\widetilde\sigma}_{dc},
\end{eqnarray}
\begin{eqnarray}\label{eqn:54}
\dot{\widetilde\sigma}_{bd}=-\gamma_{bd} {\widetilde\sigma}_{bd}
+ig_2\hat{\cal E}_2+i\Omega_2{\widetilde\sigma}_{bc} +F_{bd},
\end{eqnarray}
where $\gamma_{\mu\nu}$ are the transversal decay rates that will
be assumed $\gamma_{ba}=\gamma_{bd}=\Gamma$ in the following
derivation and $F_{\mu\nu}$ are $\delta$-correlated Langevin noise
operators. From Eqs. (\ref{eqn:52}) and {\ref{eqn:54}} we find in
the lowest (zero) order
\begin{eqnarray}\label{eqn:coherence1}
{\widetilde\sigma}_{ba}=(ig_1\hat{\cal
E}_1+i\Omega_1{\widetilde\sigma}_{bc} +F_{ba})/\Gamma,
\end{eqnarray}
\begin{eqnarray}\label{eqn:coherence2}
{\widetilde\sigma}_{bd}=(ig_2\hat{\cal
E}_2+i\Omega_2{\widetilde\sigma}_{bc} +F_{bd})/\Gamma.
\end{eqnarray}
Substitute the above two formulae into Eq. (\ref{eqn:53}) yields
\begin{eqnarray}\label{eqn:coherence3}
\dot{\widetilde\sigma}_{bc}=
\Gamma^{-1}\Omega_0^2{\widetilde\sigma}_{bc}
-\Gamma^{-1}(g_1\Omega_1\hat {\cal E}_1+g_2\Omega_2\hat {\cal
E}_2),
\end{eqnarray}
where $\Omega_0=\sqrt{\Omega_1^2+\Omega_2^2}$. The Langevin noise
terms are neglected in the above results. For our purpose we shall
calculate $\widetilde\sigma_{bc}$ to the first order, so
\begin{eqnarray}\label{eqn:coherence4}
\widetilde\sigma_{bc}\approx-\frac{1}{\Omega_0^2}(g_1\Omega_1\hat
{\cal E}_1+g_2\Omega_2\hat {\cal
E}_2)+\nonumber\\
+\frac{\Gamma}{\Omega_0^4}(g_1\Omega_1\partial_t\hat {\cal
E}_1+g_2\Omega_2\partial_t\hat {\cal E}_2).
\end{eqnarray}

According to the former discussions, here the dark- and
bright-state polaritons in the multi-mode case can be defined in
continuous form:
\begin{eqnarray}\label{eqn:DSP}
\hat\Psi(z,t)=\cos\theta(t)\hat {\cal E}_{12}(z,t) -
\sin\theta(t)\, \sqrt{N}\, \widetilde\sigma_{bc}(z,t),
\end{eqnarray}
\begin{eqnarray}\label{eqn:BSP}
\hat\Phi(z,t)=\sin\theta(t)\hat {\cal E}_{12}(z,t) +
\cos\theta(t)\, \sqrt{N}\, \widetilde\sigma_{bc}(z,t),
\end{eqnarray}
where $\hat {\cal E}_{12}(z,t)=\cos\phi(t)\, \hat {\cal
E}_1(z,t)+\sin\phi(t)\, \hat {\cal E}_2(z,t)$ is the superposition
of two quantized probe fields.

One can transform the equations of motion for the electric field
and the atomic variables into the new field variables. Similar to
the single-mode case, we consider the low-excitation approximation
and find
\begin{eqnarray}\label{eqn:DSP1}
\biggl[\frac{\partial}{\partial t} +c\cos^2\theta
\frac{\partial}{\partial z}\biggr]\, \hat\Psi(z,t)
=\dot\phi\sin\theta\cos^2\theta \, \hat
s(z,t)-\nonumber\\
-\dot\theta\, \hat\Phi(z,t) -\sin\theta \cos\theta\,
c\frac{\partial}{\partial z}\hat\Phi(z,t),
\end{eqnarray}
and
\begin{eqnarray}\label{eqn:BSP1}
\Phi&=&\frac{\Gamma}{g_1g_2\sqrt{N}}\frac{\cos^2\theta}{\Omega^2_0}\tan\theta\frac{\partial}{\partial
t}(\sin\theta\Psi-\cos\theta\Phi)\nonumber\\
&&-\sin\theta(\Omega_2^2-\Omega_1^2)\hat s(z,t),
\end{eqnarray}
where we have defined $\hat s(z,t)=-\sin\phi(t)\, \hat {\cal
E}_1(z,t)+\cos\phi(t)\, \hat {\cal E}_2(z,t)$. It is easy to see
when $\hat s=0$, the total system can be reduced to the usual
three-level one. For this we shall calculate the equation of
motion of $\hat s(z,t)$ to study the adiabatic condition. From
Eqs. (\ref{50}) and (\ref{51}) and together with the results of
${\widetilde\sigma}_{ba}$ and ${\widetilde\sigma}_{bd}$ one can
verify that
\begin{eqnarray}\label{eqn:58}
(\frac{\partial}{\partial t}+c\cos^2\beta\frac{\partial}{\partial
z})\hat
s&=&-\frac{(g_1^2\Omega_2^2+g_2^2\Omega_1^2)N}{\Gamma}\frac{\cos^2\beta}{\Omega_0^2}\hat
s-\nonumber\\
&&-\frac{1}{2}g_1g_2\sqrt{N}\sin2\beta\frac{\partial}{\partial
t}\hat {\cal E}_{12},
\end{eqnarray}
with
\begin{eqnarray}\label{eqn:mix3}
\tan^2\beta=\frac{N\Omega_1^2\Omega_2^2}{g_1^2\Omega_2^2+g_2^2\Omega_1^2}\frac{(g_1^2-g_2^2)^2}{\Omega_0^2}.
\end{eqnarray}
The time derivative of the mixing angle $\phi$ is neglected in the
above equation. The first term in the right side of Eq.
(\ref{eqn:58}) reveals a large absorption of $\hat s(z,t)$, which
causes the field $\hat s(z,t)$ to be quickly reduced to zero so
that the present system reaches pulse matching
\cite{10,match,liu}: $\hat {\cal E}_2\rightarrow\tan\phi\hat {\cal
E}_1$. For a numerical estimation, we typically set \cite{3}
$g_1\approx g_2\sim10^5s^{-1}, N\approx10^8$,
$\Gamma\approx10^8s^{-1}$, then the life time of field $\hat
s(z,t)$ is about $\Delta t\sim10^{-10}s$ which is much smaller
than the storage time \cite{3}. Furthermore, by introducing the
adiabaticity parameter $\tau=(g_1g_2\sqrt{N}T/\Gamma)^{-1}$, we
calculate the lowest order in Eq. (\ref{eqn:BSP1}) and thus obtain
$\hat\Phi\approx0, \hat s\approx0$. Then the formula
(\ref{eqn:DSP1}) reduces to the motion equation of DSPs defined in
the usual three-level $\Lambda$ type system. Consequently we have
\begin{eqnarray}\label{eqn:59}
\hat {\cal E}_1(z,t)=\cos\theta(t)\cos\phi(t)\hat \Psi(z,t)
\end{eqnarray}
\begin{eqnarray}\label{eqn:60}
\hat {\cal E}_2(z,t)=\cos\theta(t)\sin\phi(t)\hat \Psi(z,t)
\end{eqnarray}
\begin{eqnarray}\label{eqn:61}
\sqrt{N}\widetilde\sigma_{bc}(z,t)=-\sin\theta(t)\hat \Psi(z,t)
\end{eqnarray}
where $\hat\Psi$ obeys the very simple equation of motion
\begin{eqnarray}\label{eqn:62}
\biggl[\frac{\partial}{\partial t} +c\cos^2\theta
\frac{\partial}{\partial z}\biggr]\, \hat\Psi(z,t) =0
\end{eqnarray}

The above results clearly show that, for example, if the initial
condition reads $\theta\rightarrow0$ and $\phi\rightarrow0$, i.e.
initially the external control fields are much stronger
$\sqrt{g_2^2\Omega_1^2+g_1^2\Omega_2^2}\gg g_1g_2\sqrt{N}$ and
$g_2\Omega_1(0)\gg g_1\Omega_2(0)$ (the first control field is
much stronger than the second one), only ${\cal E}_1(z,t)$ is
injected into the media and the polariton $\Psi={\cal E}_1(z,t)$.
By adjusting the control fields so that
$\sqrt{g_2^2\Omega_1^2+g_1^2\Omega_2^2}\ll g_1g_2\sqrt{N}$, the
polariton evolves into
$\hat\Psi=-\sqrt{N}\widetilde\sigma_{bc}(z,t)$ and the quantum
information of the input probe pulse is stored. Likewise the
analysis in section II, when the mixing angle $\theta$ is rotated
back to $\theta=0$ again with $\phi$ to some value $\phi_e$ that
is solely determined by the Rabi-frequencies of the two reapplied
control fields, from the formulae (\ref{eqn:59}) and
(\ref{eqn:60}) one finds another quantum field $\hat {\cal
E}_2(z,t)$ will be created. The amplitudes of the two output
quantum fields are controllable by the reapplied control fields.

Now we shall give a brief discussion on the bandwidth of the probe
fields that can be stored. As an example, we will deal with the
first probe field (the discussion for another probe field is
similar). According to the results of Eq. (\ref{eqn:59}), we can
see the spectral width of the probe field narrows (broadens) when
the mixing angles change
\begin{eqnarray}\label{eqn:band1}
\Delta\omega_{p1}(t)\approx\frac{\cos^2\theta(t)\cos^2\phi(t)}{\cos^2\theta(0)
\cos^2\phi(0)}\Delta\omega_{p1}(0).
\end{eqnarray}
As in the present adiabatic condition, the propagation of the
field ${\cal E}_{12}(z,t)$ is the same with that of the probe
field in the three-level case, according to the previous results
\cite{6} we obtain its EIT transparency window to be
\begin{eqnarray}\label{eqn:band2}
\Delta\omega_{tr}(t)=\frac{\cot^2\theta(t)}{\cot^2\theta(0)}\Delta\omega_{tr}(0).
\end{eqnarray}
On the other hand, we have the relation ${\cal
E}_{1}(z,t)=\cos\phi(t){\cal E}_{12}(z,t)$, while their
wave-packet lengths keep constant during the propagation (note
that the Rabi-frequencies of control fields are independent of
space in the present case). Therefore, we can reach the
transparency window of the field ${\cal E}_{1}(z,t)$ as follows:
\begin{eqnarray}\label{eqn:band3}
\frac{\Delta\omega^{p1}_{tr}(t)}{\Delta\omega^{p1}_{tr}(0)}
\approx\frac{\cos^2\phi(t)}{\cos^2\phi(0)}\frac{\Delta\omega_{tr}(t)}{\Delta\omega_{tr}(0)}.
\end{eqnarray}
Together with the above three equations
(\ref{eqn:band1}-\ref{eqn:band3}), we can easily find
\begin{eqnarray}\label{eqn:band4}
\frac{\Delta\omega_{p1}(t)}{\Delta\omega^{p1}_{tr}(t)}
=\frac{\sin^2\phi(t)}{\sin^2\phi(0)}\frac{\Delta\omega_{p1}(0)}{\Delta\omega^{p1}_{tr}(0)}.
\end{eqnarray}
In the practical case, $\sin^2\phi(t)/\sin^2\phi(0)$ is always
close to unit. Thus absorption can be prevented as long as the
input pulse spectrum lies in the initial transparency window:
\begin{eqnarray}\label{eqn:band5}
\Delta\omega_{p1}(0)\ll\Delta\omega^{p1}_{tr}(0).
\end{eqnarray}
Obviously, this result is similar to the requirement in usual
three-level ensemble case \cite{6} and can easily be fulfilled
when an optically dense medium is used.

Finally, we shall give a brief estimate of the effect of atomic
motion. In fact, atomic motion will lead to an additional phase
evolution in the flip operators. For example, considering an atom
in position $\vec r_j$, we have
\begin{eqnarray}\label{eqn:motion1}
\hat\sigma_{bc}\rightarrow\hat\sigma_{bc}e^{i\Delta\varphi_j(\vec
r_j)},
\end{eqnarray}
where $\Delta\varphi_j(\vec r_j)=\Delta\vec k\cdot\vec r_j(t)$
with $\Delta\vec k=\vec k_{cj}-\vec k_{pj}$. Here $\vec k_{cj}$
and $\vec k_{pj}$ are wave vectors of probe and control fields and
for convenience we may assume $\vec k_{c1}-\vec k_{p1}=\vec
k_{c2}-\vec k_{p2}$. The above Eq. shows that the free motion will
result in a highly inhomogeneous phase distribution for the atoms
in different positions, and then cause the decoherence of quantum
states. In the adiabatic condition, atomic free motion can be
studied by Wiener diffusion \cite{decoherence}. According to the
results of Ref. \cite{decoherence}, the decoherence of a state
$|D_n\rangle$ is characterized by the factor $e^{-nDt}$, where $D$
is the constant diffusion rate. On the other hand, for our model
we can use co-propagating probe and control fields (see Fig. 1
(b)) so that $k_{cj}\approx\vec k_{pj} (j=1,2)$. Such a
configuration can greatly reduce the phase diffusion and then
avoid the decoherence induced by atomic free motion.

\section{conclusions}

In conclusion we present a detailed quantized description of DSP
theory in a double $\Lambda$ type four-level atomic ensemble
interacting with two quantized probe fields and two classical
control ones, focusing on the dark state evolution and the
interesting quantum memory process in this configuration. This
problem is of interest because, i) rather than one state of a
given probe light, the injected quantized field can convert into
two different output pulses by properly steering two control
fields; ii) by preparing the probe field in a non-classical state,
e.g. a macroscopic quantum superposition of coherent states, a
feasible scheme to generate optical entangled states is
theoretically revealed in this controllable linear system, which
may open up the way for DSP-based quantum information processing.
The larger class of zero-eigenvalue states besides dark-states are
identified for this system and, even in the presence of level
degeneracy, we still confirm the validity of adiabatic passage
conditions and thereby the robustness of the quantum memory
process. Furthermore, we extend the single-mode quantum memory
technique to the case with multi-mode probe fields, and reveal the
exact pulse matching phenomenon between two quantized probe pulses
in the present system. This work suggests many other interesting
ways forward, for example, by applying forward and backward
control fields in our system, we may obtain stationary pulse of
entangled states of light fields \cite{lukin}. Other issues
relation to interesting statistical phenomena such as spin
squeezing \cite{12} and possible manipulating of quantum
information \cite{13} may also comprise the subjects of future
studies.

\bigskip

\noindent We thank professors Yong-Shi Wu and J. L. Birman for
valuable discussions. We also thank Xin Liu and Min-Si Li for
helpful suggestions. This work is supported by NUS academic
research Grant No. WBS: R-144-000-071-305, and by NSF of China
under grants No.10275036 and No.10304020.




\bigskip

\noindent


\begin{thebibliography}{99}
\bibitem{1} L. V. Hau et al., Nature (London) 397, 594 (1999).
\bibitem{2} S. E. Harris, J. E. Field and A. Kasapi, Phys. Rev. A 46, R29 (1992);
            M. O. Scully and M. S. Zubairy, Quantum Optics (Cambridge University Press, Cambridge 1999).
\bibitem{3} M. M. Kash et al., Phys. Rev. Lett. 82, 529(1999);
            C. Liu, Z. Dutton, C. H. Behroozi and L. V. Hau, Nature (London) 409, 490 (2001);
            D. F. Phillips, A. Fleischhauer, A. Mair, R. L. Walsworth and M. D. Lukin, Phys. Rev. Lett. 86,783(2001).
\bibitem{4} M. D. Lukin and A. Imamoglu, Phys. Rev. Lett. 84, 1419 (2000);
            M. D. Lukin, S. F. Yelin and M. Fleischhauer, Phys. Rev. Lett. 84, 4232 (2000);
            M. Fleischhauer and S. Q. Gong, Phys. Rev. Lett. 88, 070404 (2002);
\bibitem{5} Y. Wu, J. Saldana and Y. Zhu, Phys. Rev. A 67, 013811 (2003);
            Y. Li, P. Zhang, P. Zanardi and C. P. Sun, quant-ph/0402177 (2004);
            G.Juzeli\={u}nas and P.\"{O}hberg, cond-mat/0402317 (2004);
            X. J. Liu, H. Jing and M. L. Ge, Phys. Rev. A 70, 055802 (2004).
\bibitem{wu}Y. Wu and L. Deng, Phys. Rev. Lett. 93, 143904 (2004);
            Y. Wu and X. Yang, Phys. Rev. A 70, 053818 (2004);
            Y. Wu, Phys. Rev. A, 71, 053820 (2005).
\bibitem{entangled} M. Paternostro, M. S. Kim, and B. S. Ham, Phys. Rev. A 67, 023811
                    (2003); M.D.Lukin and A.Imamo\u{g}lu, Phys. Rev. Lett, 84, 1419 (2000).
\bibitem{6} M. Fleischhauer and M. D. Lukin, Phys. Rev. Lett. 84, 5094 (2000);
            M. Fleischhauer and M. D. Lukin, Phys. Rev. A 65, 022314 (2002).
\bibitem{7} M. D. Lukin, Rev. Mod. Phys. 75, 457(2003).
\bibitem{8} C. P. Sun, Y. Li and X. F. Liu, Phys. Rev. Lett. 91, 147903 (2003).
\bibitem{9} A. B. Matsko, et al., At. Mol. Opt. Phys. 46, 191 (2001);
            A. S. Zibrov et al., Phys. Rev. Lett. 88, 103601 (2002).
\bibitem{10} A. Raczy\'{n}ski and J. Zaremba, Opt. Commun. 209, 149 (2002); quant-ph/0307223 (2003).
\bibitem{entanglement} O. Hirota, quant-ph/0101096(2001).
\bibitem{ent} D. N. Matsukevich and A. Kuzmich, Science, 306, 663 (2004);
              C. H. van der Wal et al., Science, 301, 196 (2003); O. Mandel et al., Science, 425, 937 (2003);
              K. Hammerer et al., arXiv: quant-ph/0312156 (2003).
\bibitem{swap} Xiaoguang Wang and Barry C. Sanders, Phys. Rev. A 65, 012303 (2003);
               F. L. Kien et al., Phys. Rev. A 68, 063803 (2003);
               N. A. Ansari et al., Phys. Rev. A 50, 1492 (1994).
\bibitem{11} C. P. Sun, Phys. Rev. D {\bf 41}, 1318 (1990);
             A. Zee, Phys. Rev. A {\bf 38}, 1 (1988).
\bibitem{match} S. E. Harris, Phys. Rev. Lett. 70, 552 (1993).
\bibitem{liu}  X. J. Liu, H. Jing, X. T. Zhou and M. L. Ge, Phys. Rev. A, 70, 015603 (2004);
               X. J. Liu, H. Jing and M. L. Ge, Chin. Phys. Lett. 23, 1184-1187 (2006).
\bibitem{decoherence} C. Mewes and M. Fleischhauer, Phys. Rev. A 72, 022327 (2005); C. W. Gardiner, \textit{Handbook of
Stochastic Methods}, Springer, Berlin, 1983.
\bibitem{lukin} M. ajcsy, A.S. Zibrov and M. D. Lukin, Nature(London), 426,638(2003).
\bibitem{12} A. Andr\'{e}, L. M. Duan and M. D. Lukin, Phys. Rev. Lett. {\bf 88}, 243602 (2002);
             L. M. Kuang and L. Zhou, Phys. Rev. A 68, 043606 (2003);
             A. Dantan and M. Pinard, quant-ph/0312189 (2003).
\bibitem{13} R. G. Beausoleil, W. J. Munro, D. A. Rodrigues and T. P. Spiller, quant-ph/0403028 (2004).

\end{thebibliography}
\end{document}